# Chiral pair density waves with residual Fermi arcs in RbV$_3$Sb$_5$

Xiao-Yu Yan,[1,*] Hanbin Deng,[1,*] Tianyu Yang,[1,*] Guowei Liu,[1] Wei Song,[1] Hu Miao,[2] Hechang Lei,[3,4] Shuo Wang,[5,6] Ben-Chuan Lin,[5,6] Hailang Qin,[7,†] Jia-Xin Yin[1,7,†]

[1]Department of Physics, Southern University of Science and Technology, Shenzhen 518055, Guangdong, China.
[2]Materials Science and Technology Division, Oak Ridge National Laboratory, Oak Ridge, Tennessee 37831, USA.
[3]Beijing Key Laboratory of Optoelectronic Functional Materials MicroNano Devices, Department of Physics, Renmin University of China, Beijing 100872, China.
[4]Key Laboratory of Quantum State Construction and Manipulation (Ministry of Education), Renmin University of China, Beijing, 100872, China.
[5]Shenzhen Institute for Quantum Science and Engineering, Southern University of Science and Technology, Shenzhen, 518055, China.
[6]Guangdong Provincial Key Laboratory of Quantum Science and Engineering, Southern University of Science and Technology, Shenzhen, 518055, China.
[7]Quantum Science Center of Guangdong-Hong Kong-Macao Greater Bay Area, Shenzhen 518055, Guangdong, China

*These authors contributed equally to this work.
†Corresponding authors. E-mail: yinjx@sustech.edu.cn; qinhailang@quantumsc.cn;

**The chiral 2×2 charge order has been reported and confirmed in the kagome superconductor RbV$_3$Sb$_5$, while its interplay with superconductivity remains elusive owing to its lowest superconducting transition temperature T$_C$ of about 0.85K among the AV$_3$Sb$_5$ family (A=K, Rb, Cs) that severely challenges electronic spectroscopic probes. Here, utilizing dilution-refrigerator-based scanning tunneling microscopy (STM) down to 30mK, we observe chiral 2×2 pair density waves with residual Fermi arcs in RbV$_3$Sb$_5$. We find a superconducting gap of 150μeV with substantial residual in-gap states. The spatial distribution of this gap exhibits chiral 2×2 modulations, signaling a chiral pair density wave (PDW). Our quasi-particle interference imaging of the zero-energy residual states further reveals arc-like patterns. We discuss the relation of the gap modulations with the residual Fermi arcs under the space-momentum correspondence between PDW and Bogoliubov Fermi states.**

RbV$_3$Sb$_5$ is the last material found to be superconducting in the AV$_3$Sb$_5$ family and it features the lowest T$_C$ ~ 0.85K (based on the middle point of the resistance-temperature curve) [1,2]. Its superconducting feature is thus much less reported and has also been suggested to be unusual [3-7]: muon spin resonance experiment reveals time-reversal symmetry-breaking and strong deviation from an isotropic pairing state [5]. Transport experiment on thin flakes reveals rotational symmetry-breaking and hysteresis of the magnetoresistance regarding the superconducting transition [6,7]. Therefore, a high-resolution electronic spectroscopic characterization of its superconducting properties is highly desirable and will advance our understanding of the emergent superconductivity in the kagome lattice in general. In this work, we perform a systematic STM experiment on single crystals of RbV$_3$Sb$_5$ down to 30mK to uncover its superconducting properties. We cleave the single crystals of RbV$_3$Sb$_5$ at 77K and focus our research on the Sb layer tightly bound to the V kagome lattice. Our STM data is mainly taken at the lattice temperature of 30mK and the electronic temperature of our instrument is evaluated to be 90mK based on the fitting of the full superconducting gap structure in $Cs(V_{0.86}Ta_{0.14})_3Sb_5$ with BCS gap function [8], unless otherwise specified. To reveal the tiny electronic variations, the tunneling junction is set at extreme conditions at a bias voltage of 0.5mV, a tunneling current of 1nA, and a modulation voltage of 5μeV for the gap map and 40 μeV for the quasi-particle interference map.

Figure 1(a) shows its crystal structure which consists of Rb layers and $V_3Sb_5$ trilayers. Figures 1(b) and (c) show the atomically resolved topographic image. The Fourier transform of the topographic image reveals extra modulations beyond the Bragg peaks [Fig. 1(d)], which include the 2×2 charge order and 1×4 stripes. The 2×2 charge order has been confirmed by bulk X-ray scattering [9], while the 1×4 stripes have not been detected by bulk scattering techniques and can be of surface origin [9]. We find that the circular quasi-particle interference signal also exhibits anisotropy, which may result from the combined anisotropy from both orders. The low energy differential conductance data in Fig. 1(e) reveals a gap-like feature with a pair of peaks located at ±150μeV, as well as substantial residual in-gap states. This feature disappears above its $T_C$, and can represent its superconducting gap. Crucially, in Fig. 1(f) we plot the determined superconducting gap size Δ (measured by the coherent peak-to-peak distance) versus $T_C$ for all three $AV_3Sb_5$ and a Ta-doped $CsV_3Sb_5$ crystal measured by the same instrument [8]. They all roughly fall into the same line predicted by the BCS theory with $2\Delta/k_BT_C=3.53$, where $k_B$ is the Boltzmann constant. This scaling relation reconfirms our identification of the superconducting gap feature in $RbV_3Sb_5$.

After identifying the superconducting gap, we further measure its spatial distribution with ultra-high precision. We use the in-plane walker to shift the target map area to the scanning zone center and scan the area for days before performing the gap map, which reduces the spatial drift of the map. We measure the superconducting gap $\Delta(r)$ by reading out the peak-to-peak distance of the spectrum taken at each spatial location *r* for a large field of view as shown in Figs. 2(a) and (b). The spatial distribution of $\Delta(r)$ in Fig. 2(b) suggests that the gap variation is on the order of 10μeV. This gap variation is further elucidated in Fig. 2(c) by two representative differential conductance spectrums and the differential conductance spectrums in Fig. 2(d) taken along the line marked in Fig. 2(a). Here, we comment on the energy resolution of STM. The commonly discussed energy resolution dE can be understood as that if an electronic state features δ function, then it will behave as a Gaussian-like function in the STM data with a full width at half height of dE, where in our case $dE=90mK·3.5k_B$. At a fixed position, if there are two such states with energy separation less than dE, then these two states will merge into one peak in the tunneling data and we cannot directly distinguish the two states. However, if these two states are at different locations, even though these two states are broadened, their spectral peak energy difference at different locations may still be distinguished by STM. In the latter case, the spatial resolution of the STM tip and the signal-to-noise ratio of the tunneling spectrum become equally crucial as the electronic energy resolution dE, which can be achieved by preparing a stable tunneling condition and elongating the measuring time. In the current case, we find that STM can resolve the spatial modulations of superconducting coherence peaks on 1μeV level [8].

We perform the Fourier transform of the gap map in Fig. 2(a) to obtain Δ(*q*) in momentum space in Fig. 2(e). We find that the superconducting gap exhibits 2×2 modulations. The 2×2 charge order in $RbV_3Sb_5$ has been reported to be chiral [10,11], where the three pairs of 2×2 vector peaks in the *q* space have different intensities and the counting direction from low to high defines clockwise or anti-clockwise chirality. In the Δ(*q*) map, we find the 2×2 vector peaks also have different intensities as marked in Fig. 2(e), and the counting direction from low to high defines a clockwise chirality. This is further illustrated in the three-dimensional plot in Fig. 2(f). To elaborate its chirality in real space, we perform an inverse Fourier transform of the 2×2 vector peaks in Fig. 2(g). Then we extract the linecut profile along three *a*-axis directions, showing that the gap modulation strength is quite different. These data on the pairing gap modulations thus demonstrate a chiral 2×2 PDW state [12-15]. In $KV_3Sb_5$, we have used magnetic field history to control the chirality of 2×2 PDW [8]. However, in the case of $RbV_3Sb_5$, after applying the magnetic field of 1T, we find that the superconducting coherence peak is substantially suppressed, making it ambiguous to accurately obtain the gap map, and the field control experiment may require an even lower temperature that goes beyond our current instrument. Crucially, together with the high precision gap map analysis in $KV_3Sb_5$ and $CsV_3Sb_5$ [8], our identification of the chiral 2×2 PDW in all three $AV_3Sb_5$ strongly constrains discussions of PDW in kagome lattice [16-19]. We also comment on the tiny modulations of the pairing gap. We find in the calculation that the pairing gap modulation is generally much smaller than the modulation of the superconducting order parameter, and they are comparable only when the system is at the perfect nesting condition. Thus the measured gap modulations can often underestimate the strength of PDW order.

Finally, we investigate the nature of the substantial residual in-gap states by performing the quasi-particle interference experiment [20-22] at zero energy at 0.9K [Fig. 3(b)] and 30mK [Fig. 3(c)] for the same area shown in Fig. 3(a). Their Fourier transform data g($q$) are shown in the insets of Fig. 3(b) and (c), respectively. We further perform six-fold symmetrization to enhance the signal-to-noise ratio of g($q$) in Fig. 3(d). A direct subtraction of g($q$) data at 0.9K and that at 30mK reveals primarily a circular signal [Fig. 3(e)], which mainly arises from the scattering of the circular pocket of the Sb *p* orbital [23]. This observation is consistent with that in $KV_3Sb_5$ [8], suggesting that the uniform superconductivity mainly occurs at the Sb *p* orbital. In the inset of Fig. 3(c), it is clear that the residual interference features at 30mK are mostly along the real space stripe direction. Therefore, we further perform mirror symmetrization to enhance the signal-to-noise ratio to highlight this observation in Fig. 3(f), where the nematic arc-like interference patterns $q_{d1}$ and $q_{d2}$ are observed. These residual interference patterns resemble that observed in $KV_3Sb_5$ [8], except that they are now mainly observed in one direction.

We discuss the possible mechanism for the observation of these residual Fermi arc states. Following the discussion of the residual Bogoliubov Fermi arcs in $KV_3Sb_5$ [8], the uniform pairing mainly occurs in the Sb *p* orbital, and the pairing is strongly suppressed in the V *d* orbital likely owing to the time-reversal symmetry-breaking chiral 2×2 charge order developed therein [24-27]. The 2×2 charge order reconstructs the Fermi surface, and an interorbital *p-d* 2×2 PDW will create Bogoliubov Fermi arcs as shown in Figs. 3(g) and (h). Lastly, the 1×4 stripes may lead to extra nematicity that the Bogoliubov Fermi arcs are stronger in one direction, and related vector $Q_{1×4}$ may connect two ends of the longer arcs leading to shortening of $q_{d1}$ arcs. Eventually, the intraband scattering of the residual Fermi arcs may give rise to the observed interference signal.

In short, the detection of chiral 2×2 PDW and residual Fermi arcs in all three $AV_3Sb_5$ points to a ubiquitous unconventional superconducting behavior arising from the intertwined order. Our observation of chiral PDW at the bulk 2×2 ordering vectors is also in line with the transport detection of the anisotropic Josephson effect [28], charge 6e superconductivity [29], superconducting diode effect [30], and the hysteresis of magnetoresistance [7], all requiring breaking composite symmetries in the superconducting state. It has not escaped our attention that large residual in-gap states are also detected in other time-reversal symmetry-breaking superconductor candidates, including $Sr_2RuO_4$ and $UTe_2$ [31,32], and it would be interesting to investigate the nature of the residual in-gap states therein with similar methodology. We look forward to the spectroscopic demonstration of the tentative topological feature of the chiral PDW in future. For instance, there can be multi-domains [33,8] of the charge order and intertwined PDW, and it would be interesting to check whether there is an additional emergent in-gap state at the domain boundaries arising from the bulk-boundary correspondence of the tentative topological PDW. Possible transverse thermal conductivity can be searched for in kagome superconductors as another piece of evidence for the topological nature of the chiral PDW.

**Figures:**

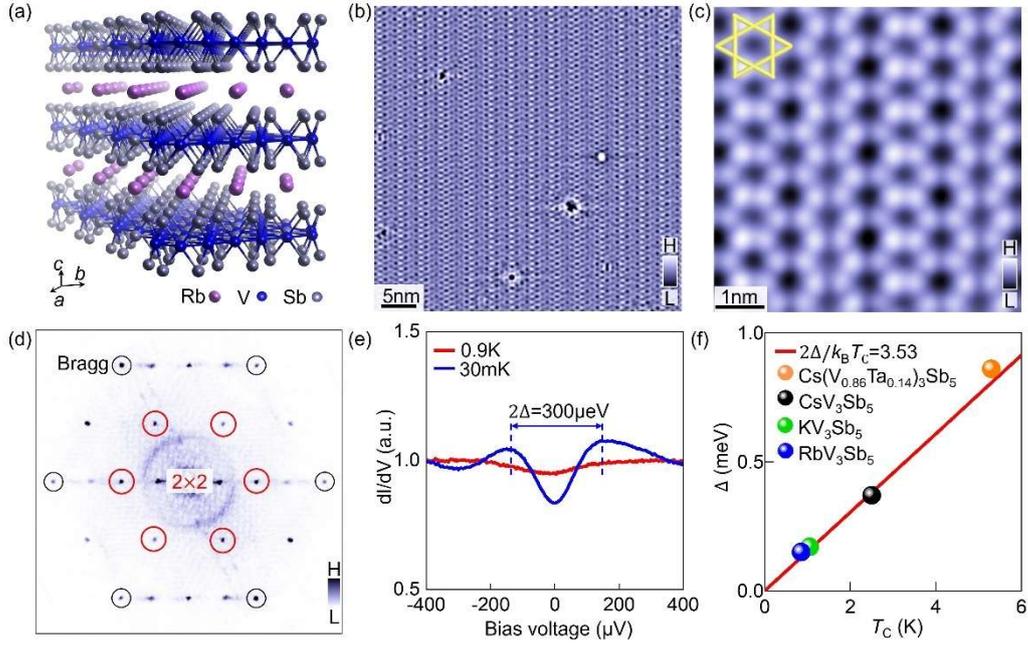

Fig. 1 Superconducting gap in RbV$_3$Sb$_5$. (a) Crystal structure of RbV$_3$Sb$_5$. (b) Topographic image of a large Sb terminating surface. (c) Zoom-in image of the Sb surface and the yellow lines mark the underlying kagome lattice. (d) Fourier transform of the large topographic image, showing Bragg peaks marked by black circles and 2×2 vector peaks marked by red circles. (e) Differential conductance spectrums taken at 30mK (well below superconducting transition temperature $T_C$) and 0.9K (slightly above $T_C$). The 30mK data exhibits a superconducting gap Δ of 150μeV. (f) Δ versus $T_C$ for all three AV$_3$Sb$_5$ (A=K, Rb, Cs) and Ta-doped CsV$_3$Sb$_5$. The red line marks the BCS scaling relation.

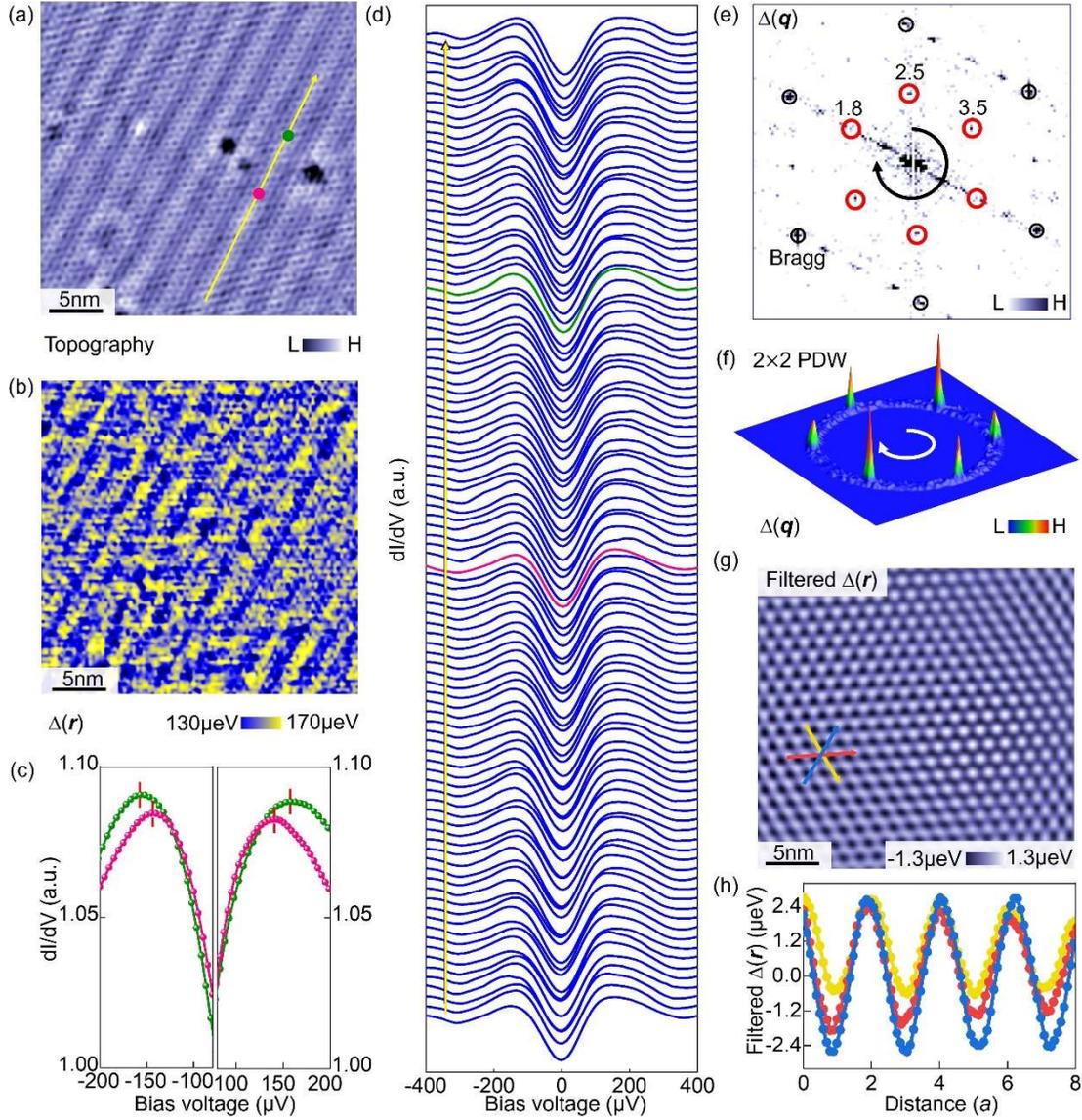

Fig. 2 Chiral 2×2 PDW order. (a) Topographic image of Sb surface. (b) superconducting gap map obtained in the area shown in (a). (c) Coherence peak data for two locations marked in (a), showing the energy gap difference. (d) Differential conductance data taken along a line marked in (a). (e) Fourier transform of the superconducting gap map, showing 2×2 vector peaks with different intensities. Counting from low to high intensities, we define the chirality to be clockwise. The black dots mark the Bragg peaks and the red circles mark the 2×2 vector peaks. (f) A three-dimensional plot of the 2×2 vector peaks. (g) Inverse Fourier transform of the 2×2 vector peaks. (h) Line profiles along three $a$-axis directions marked in (g), showing different modulation strengths, elaborating the chirality in real space.

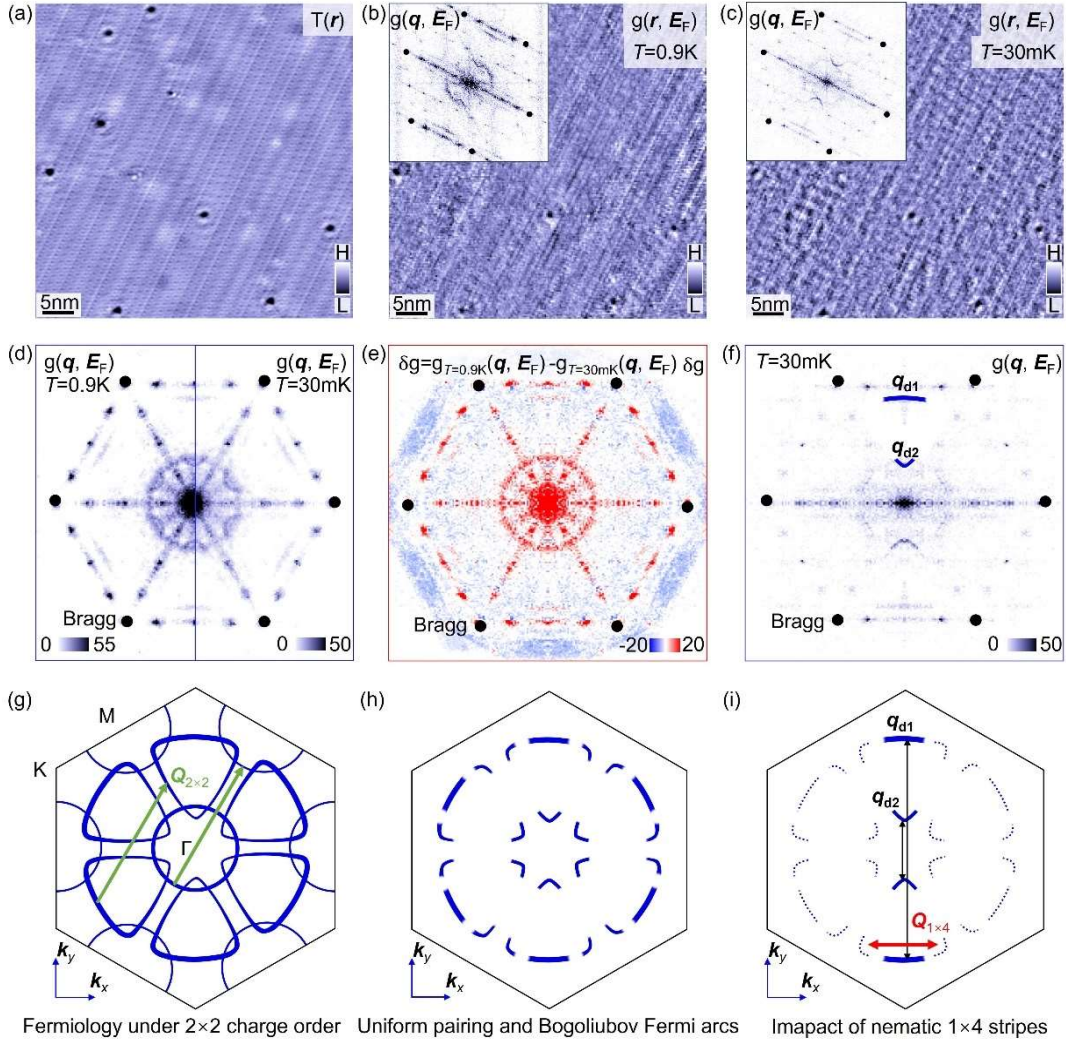

Fig. 3 Residual Fermi arcs and their relation with PDW. (a) Topographic image of Sb surface. (b) Corresponding differential conductance map taken at zero-energy at 0.9K. The inset shows its Fourier transform. (c) Corresponding differential conductance map taken at zero-energy at 30mK. The inset shows its Fourier transform. (d) Six-fold symmetrized zero-energy quasi-particle interference data at 0.9K (left) and 30mK (right). (e) Subtraction of the interference data at 0.9K and that at 30mK. (f) Mirror symmetrized zero-energy interference data at 30mK, showing residual Fermi arcs at $q_{d1}$ and $q_{d2}$ marked by blue curves. (g) Fermiology under the 2×2 charge order, showing circular pockets of Sb $p$ orbital and triangular pockets of V $d$ orbital. (h) Uniform pairing gaps out the circular pocket and a $p$-$d$ PDW further gaps out the intersections between $p$ and $d$ orbitals, leaving Bogoliubov Fermi arcs. (i) The 1×4 stripes further lead to nematicity. The intraband scattering of the remaining Fermi arcs may lead to the observed signals at $q_{d1}$ and $q_{d2}$.

**Acknowledgement:**
We acknowledge the support from the National Key R&D Program of China (Nos. 2023YFA1407300, 2023YFF0718403), the National Science Foundation of China (Nos. 12374060, 12074162), Guangdong Provincial Quantum Science Strategic Initiative (No. GDZX2201001), and Guangdong Basic and Applied Basic Research Foundation (No. 2022B1515130005).